\documentclass[12pt]{article}

\usepackage{scicite}

\usepackage{times}
\usepackage{graphicx}

\newenvironment{sciabstract}{%
\begin{quote} \bf}
{\end{quote}}

\newcounter{lastnote}

\title{Magnetars From  Magnetized Cores Created by a  Strong Interaction Phase Transition}

\author
{N.D. Hari Dass,$^{1\dag}$ Vikram Soni,$^{2\ast}$ \\
\\
\normalsize{$^{1}$
CHEP, Indian Institute of Science, Bangalore 560012, INDIA }\\
\normalsize{$^{2}$
Centre for Theoretical Physics, Jamia Millia University, New Delhi 110025, INDIA }\\
\\
\normalsize{$^\dag$E-mail:  dass@cts.iisc.ernet.in}\\
\normalsize{$^\ast$
E-mail:  v.soni@airtelmail.in}
}

\date{}

\begin{document} 


\baselineskip24pt


\maketitle


\begin{sciabstract}We consider a model where the strong magnetic fields of magnetars arise from
 a high baryon density, magnetized core. In this framework magnetars are distinguished
from pulsars by their higher masses and central density. For magnetars, as core
densities exceed a threshold, the strong interaction induces a phase transition
to a ground state
%a neutral pion condensate
 that aligns all magnetic moments.
The core magnetic field is initially shielded by the ambient high
conductivity  plasma. With time the shielding currents dissipate
transporting the core field out, first to the crust and then breaking through the
crust to the surface of the star. Recent observations provide strong
 support for this model which accounts for several properties of magnetars and also
 enables us to identify new magnetars.

  %This document presents a number of hints about how to set up your
  %{\it Science\/} paper in \LaTeX\ .  We provide a template file,
  %\texttt{scifile.tex}, that you can use to set up the \LaTeX\ source
  %for your article.  An example of the style is the special
  %\texttt{\{sciabstract\}} environment used to set up the abstract you
  %see here.
\end{sciabstract}

%%%%%%%%%%%%%%%%%%%%%science.doc read in %%%%%%%%%%%%%%%%%%%%%%%

% In setting up this template for *Science* papers, we've used both
% the \section* command and the \paragraph* command for topical
% divisions.  Which you use will of course depend on the type of paper
% you're writing.  Review Articles tend to have displayed headings, for
% which \section* is more appropriate; Research Articles, when they have
% formal topical divisions at all, tend to signal them with bold text
% that runs into the paragraph, for which \paragraph* is the right
% choice.  Either way, use the asterisk (*) modifier, as shown, to

\section*{Introduction}
%%%%introhere%%%%%%%%%%%%%%%%%%%%%%%%
%\section{Introduction}
{\label{intro}}
Magnetars are neutron stars with surface magnetic fields ( $10^{14(15)}$ G) . 
The magnetars have spin down ages of $10^{3}-10^{5}$ years. Over this period, they emit a quiescent radiative
luminosity of $10^{35}$ - $10^{36}$ erg/s. Besides, some of them emit repeated flares or bursts
of energy typically of $10^{42} - 10^{44}$ erg, and at times much higher 
\cite{palmer+05,hurley+05}. The periods of magnetars fall in a surprisingly narrow window of 5-12 s.

At such large periods, the energy emitted in both quiescent emission and flares far exceeds the loss in their rotational energy . The most likely energy source for these emissions is their magnetic energy \cite{duncan+thompson92,thompson+duncan95}, yet there is no evidence of a decrease in their surface magnetic fields with time \cite{thompson+02}.

There have been many attempts to explain some of this physics of which the most canonical  is the magnetar model of Duncan and Campbell \cite{duncan+thompson92,thompson+duncan95}, which is otherwise known as the {\em dynamo mechanism for magnetars}. This model requires the collapse of a large mass progenitor to a star which starts life with a period close to a millisecond. Such a fast rotation can amplify the inherited pulsar valued field of $10^{12}$
G to $10^{15}$ G. However, as described below, several observations on magnetars are hard to understand from such a  model.

 Many authors, Leahy et al \cite{leahy0}, Wickramsinghe et al \cite{wickramasinghe}, Gaensler et al \cite{gaensler0}
and Muno \cite{muno0}, have observed that the magnetars are descendents of large mass progenitors or protostars. 
Wickramsinghe et al \cite{wickramasinghe}, based on simulations, further suggest that magnetars may actually belong to the higher ( than pulsars) mass ($\simeq 1.6 M_{sun}$) population of neutron stars.This in itself would be consistent with the dynamo model as stars descended from large mass progenitors have the smallest periods. 
However, as deduced by Vink et al \cite{vink0}, from the explosion energies of the Super Nova remnants (SNR’s) of
magnetars and Heger et al \cite{heger0}, using stellar evolution codes, even such massive stars  will be born with
periods  ($\simeq 5 ms$ ) somewhat larger than what is required for dynamo amplification.

The spin down age of magnetars deduced from their pulsar like dipole radiation is not
the real age for the magnetars, which is instead given by the age of their supernova
remnants(SNR) and is much larger ( see \cite{ddarxiv,ddgreco}and \cite{leahy0}). 
{Leahy et al \cite{leahy0} point out that this can} be explained if it takes 
an additional time of the order of $10^4$ yrs for the magnetar surface magnetic field 
to attain its high final value.  They interpreted it as a {\em delayed amplification}. 
Such a delayed amplification would not conform to the dynamo model.

If magnetars are born with thier high magnetic fields,then, after a flare, one would expect a {\em decrease} in magnetic field and energy of the magnetar {accompanied by }a fall in the dipole radiation mediated spin down rate. However, the opposite is seen: – the spin down rate {\em increases} after the flare indicating an increase in the surface magnetic field
( see \cite{ddarxiv,ddgreco,dar1999a,marsden0}). Further, inspite of the steady X-ray emission, the magnetic field of such magnetars appears to remain high all the way till the end of spin down, when the stars have the largest periods.

{These striking discrepancies with the dynamo model }indicate that we need an alternative model for magnetars.

In an earlier work \cite{dbvsarxiv} it was shown that it may be  possible to explain
many unusual features of magnetars if they have a core with a large magnetic moment density, created by the strong interaction. Initially  the core magnetic field is shielded by the electron plasma in and around the core. In time, the shielding currents dissipate till finally the field emerges at the surface of the star.

In this work we break new ground. In section II we give a new two stage
 description of the emergence of the shielded core field from the core to the crust
and then from the crust to the surface.This includes a
consistent energy budget to set up the shielding currents and their
subsequent decay into neutrinos, quiescent radiation and flares.
Section III is new and devoted to detailing the extended observational  support
for the model. In section IV we use this model to establish new criteria  which enable
us to identify new magnetars in a list of several high magnetic field pulsars, whose
magnetic fields are smaller than assumed for magnetars ( $10^{14(15)} G$) or periods
much less than typical magnetars periods.

\section*{The Model}
Pulsars, which have radii of $ \simeq 10 km $ and surface magnetic fields of $ \simeq 10^{10}- 10^{12}$ G, are believed to inherit such fields from their progenitors, which are stars of radius $ \simeq 10^{6} km $ and  magnetic fields of $ \simeq 1 - 10^{2}$ G  due to 'conservation' of magnetic flux during stellar collapse \cite{woltjer64}.
We call such fields {\em inherited}  (some authors call them {\em fossil} fields).

Our starting point, to make the distinction between pulsars and magnetars, is that most observed
pulsars have masses of the order of 1.4 solar masses. 
Theory allows pulsar(neutron star) masses
in the larger range of 0.1 - 2.0 solar masses. 
Pulsars with masses larger than 1.4 solar masses
%have definitely been seen 
are also known \cite{pulsarmasses}. Magnetars are expected to be larger mass stars, whose core density exceeds about three times the  nuclear density. This leads to a new strong interaction ground state -a neutral pion condensate - which aligns the spins of the neutrons (quarks) in the core producing  a very large dynamical (spin) magnetization of the core, which can give  core magnetic fields as large as $10^{16} - 10^{17} $G \cite{dbvsplb,dbvshepph}.

As the strong interaction phase transition starts aligning the magnetic moments of the neutrons(quarks) the electron and proton plasma in the core responds  to the local change of flux, shielding the magnetic field in accordance with Lenz's law. This process goes on till the phase transition in the core is completed and all magnetic moments are aligned.The shielding currents in and around the core confine the magnetic field inside a volume of the order of the core. The typical energy in the shielding currents is theN of the order of the magnetic energy of the core. Eventually, the Lenz currents dissipate establishing the full unshielded dipolar field in and outside the core .
 In the process of this dissipation there is a characterstic time during which ambipolar diffusion \cite{goldreisenegger92,spitzer78} carries the core field to the crust. The magnetic field then breaks out of  the crust to power the radiative emissions from magnetars.

%\subsection{ Creation Of The Core}

{\bf Creation Of The Core}:
%As shown by us such a 
%$\pi_0$ condensate aligns the magnetic moments of neutrons in nuclear medium and of quarks in
%quark matter, leading to huge core magnetizations.
% The energy reduction from spin
%alignment comes from the condensate which introduces an extra term −GA/2~q · ~3 in the
%strong interaction Hamiltonian where ~q is the condensate wave vector, ~ the spin operator of
%the nucleon and 3 the third component of the isospin. This e0 condensate state aligns all the
%neutron (see (22)) ( or quark ( see (23)) ) spins, and thus magnetic moments, to achieve the
%lowest energy.
%For a star composed entirely of neutrons with an average density of 5 times nuclear density
In our model magnetars belong to the higher mass population of neutron stars. They are 
distinguished by high density cores which go through a  phase transition to a spin aligned 
ground state that carries a large magnetic moment. One likely possibility is a spin aligned
 neutral pion condensate. Such a high density ground state was first 
proposed by Dautry and Nyman \cite{dautry}. A simple way to understand how such a 
’magnetic’ state can occur, for example, in a star made of neutrons is as follows. The pion 
condensate standing wave introduces an extra term $-{g_A}{\vec q}\cdot{\vec\sigma}\tau_3/{2}$ 
in the in the strong interaction Hamiltonian ( chiral sigma model ) for nucleons,
where ${\vec q}$ is the condensate wave vector, ${\vec \sigma}$ the spin operator of the nucleon 
(quasiparticle) and $\tau_3$ the third component of the isospin. For neutrons the third component 
of the isospin is negative, and thus, if the spins of the neutron quasiparticles align
antiparallel to the direction of the condensate wave vector, the energy is considerably reduced.

 Baym \cite{baym77}, has a simple exposition on the neutral $\pi_0$ condensed state in the 
chiral sigma model based on the work of Dautry and Nyman \cite{dautry}. The ground state energy 
for the aligned state is given by summing over all (one spin) states till the fermi momentum. The 
minimization of the ground state energy with respect to the condensate wave vector, $q$, yields a, 
$q$, that is proportional to the baryon density. The ground state energy gets a term from  the  
condensate that is negative and proportional to, $ n^2$,where $n$ is the neutron number density. 
On the other hand, the kinteic energy term for the nucleons goes as, $ n^{5/3}$. As the density 
goes up the negative condensate term dominates leading to a phase transition to the neutral 
$\pi_0$ pion condensed groundstate for large baryon density. However, inclusion of other 
nucleon-nucleon interactions like the tensor interaction and the hardcore may oppose 
such a spin alignment of the $\pi_0$ condensed ground state, whereas isobars may enhance it. Further 
work is needed to find a conclusive answer.
 
 The same phase transition to the neutral $\pi_0$ pion condensed ground state occurs for the 
equation of state (EOS) for two flavour quark matter\cite{kutschera+90}. However, as the quarks 
are taken to be point particles, in this case we do not have tensor and hard core forces that 
can undo the spin alignment of the $\pi_0$ condensed ground state. 
This ground state has been has been analyzed in  \cite{kutschera+90}, 
who use a ground state theorem to find substantial magnetization for the $\pi_0$ condensed ground 
state ( see eqn. 50 - 52 in ref\cite{kutschera+90}).

In previous papers \cite{dbvsarxiv,dbvsplb,dbvshepph} we investigated the ground
state
%s of both purely nuclear matter as well 
OF composite quark and nuclear matter stars.
% We had
%found that in both cases, beyond a threshold density the ground state of strongly interacting
%matter is likely to be a standing wave neutral $\pi_0$ condensate. The 3 flavour quark matter has been analysed 
%in \cite{dbvsprd}.
 The equation of state (EOS) and magnetic dipole moment density for a neutron star 
with quark matter core having a neutral $\pi_0$ pion condensed ground state is calculated 
in \cite{dbvsplb}. The energy per baryon for the neutral $\pi_0$ pion condensed groundstate and the usual fermi 
sea (the uniform, $ q = 0$ state) is plotted in fig 1(b) in \cite{dbvsplb}). This is reproduced 
in Figure. 1, from which we can read out the energy released in the core phase transition to 
the neutral $\pi_0$ pion condensed  state to be of the order of ~ 30 Mev
($4.5\cdot 10^{-5}$ ergs)  per baryon. 

\begin{figure}[h!]
\begin{centering}
\includegraphics[scale=0.30]{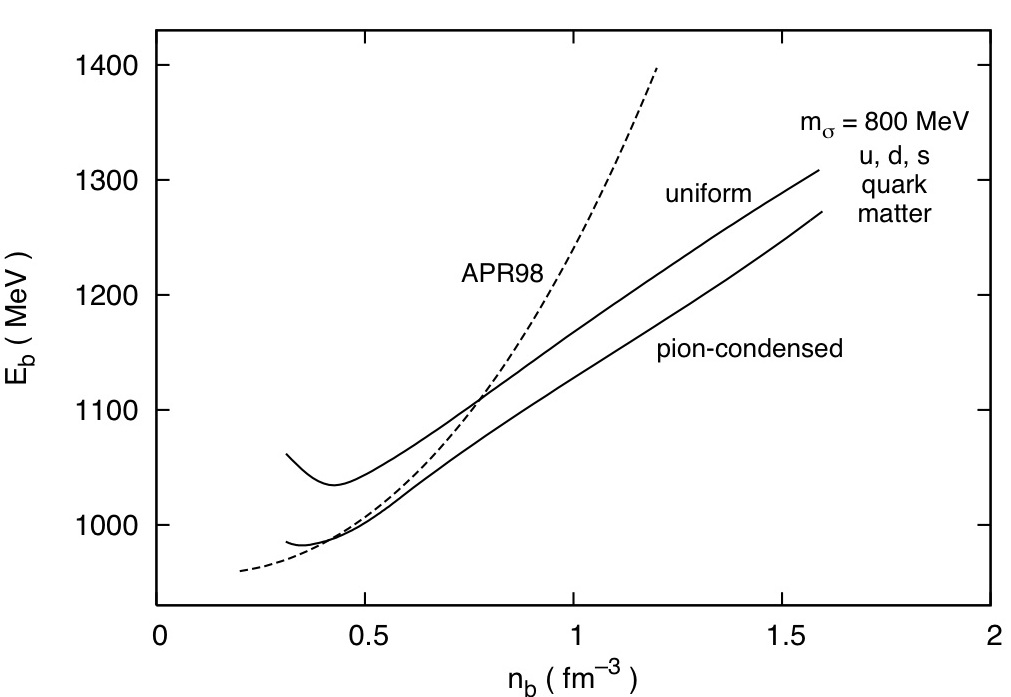}
\caption{The Energetics of The Neutral Pion Condensate Transition(From Bhattacharya and Soni \cite{dbvsplb})}
\end{centering}
\end{figure}

We note that a quark matter core  occurs \cite{dbvsplb} only if the nuclear matter –quark matter 
occurs at low pressure –otherwise the higher pressure nuclear exterior will squeeze out the softer
, relativistic quark interior.
Such conditions obtain if the minimum in  the energy per baryon, $E_{BN}$ vs $ n_B$ , the baryon 
density, in the nuclear phase is close to the minimum in the energy per baryon, in the quark
matter phase, $E_{BQ}$ vs $ n_B$; for then, the slope of the tangent to the two minima in the, 
$E_B$ vs ( $1/n_B$)
diagram ( the so called Maxwell construction ), which gives the pressure, is small.

For illustration  we consider a star composed entirely of neutrons with an average core density 
of ~5 times nuclear density  ($\simeq 10^{15} gcm^{-3}$).In a unit volume there are some $\simeq  
10^{39}$ neutrons each with a magnetic moment of $\mu_N \simeq 2\cdot{e\hbar}/{2 m_N c} 
\simeq 10^{-23}$ (CGS units) giving a magnetic moment density of $m \simeq  10^{16}$ (CGS units). 
This would result in  a uniform core field of $B={8\pi}/{3}\cdot m \simeq 5\cdot 10^{16}$ G. 
With the reasonable value of 3 kms for the core radius, this would give a surface magnetic
field of $\simeq 10^{15}$ G. A similar result follows {for} a composite star which has a 
quark-matter core \cite{dbvsarxiv}.
Unlike magnetic fields whose origins are currents, this field built out of strong interaction 
mediated alignment of magnetic moments is {\em indestructible} for all practical purposes.

%\subsection{Magnetic Field Evolution In Magnetars}
%\subsubsection{The First Stage}

{\bf Magnetic Field Evolution - The First Stage}:
The ground state of the core, above, is the most likely (variational) possibility without any 
clinching proof. For the rest of the paper we shall assume that the strong interaction creates 
a magnetized core and then discuss the evolution of the magnetic field from such a core by the 
dissipation of shielding currents.
Time scales for these processes, relevant to a neutron, proton, electron plasma in the interior 
of a neutron star,
have been worked out by Goldreich and Reisenegger \cite{goldreisenegger92}. Their estimates 
show that
for typical temperatures in the neutron star interior, the ohmic dissipation time scale would be 
very long: $> 10^{11}$ y, and hence irrelevant. Ambipolar diffusion will, however, play a very 
important role, with a dissipation time scale of  $10^4 \cdot B_{16}^{-2} \cdot T_{8.5}^{-6}$ 
years, where $B_{16}$ is the local magnetic field strength in units of $10^{16}$ G and $T_{8.5}$ 
is the temperature in units of $10^{8.5}$ K, a typical value in the interior
of a very young neutron star.

Ambipolar diffusion lets the magnetic field move out from the core to the crust. The high 
conductivity interior plasma in the star is opaque to photons and only the neutrinos are 
likey to escape \cite{kaminker}. The temperature must 
be higher than $\simeq 10^{8.5}$ K to have neutrino emission as the dominant energy release 
mechanism. The photon emission which is trapped will keep heating the star interior to this 
temperature but neither the luminosity of the star or the surface magnetic field will change 
till the magnetic field crosses out of the crust to the surface. 
At temperatures $>10^{8.5}$ K and fields of $10^{ 16(15)}$ G in the interior the ambipolar diffusion 
formula gives a typical travel time of $10^{4(5)}$  years from core to crust. This phenomenon 
(absent for pulsars) will manifest as a delayed amplification of the magnetar activity and 
will provide an observable window to the core. It is important to emphasize that ambipolar
diffusion which carries the fields outwards is also accompanied by dissipation of magnetic energy.
A fine balance is necessary between the rate of this dissipation and the rate of neutrino cooling
to keep these temperatures for as long as $10^4$ yrs. This would also result in a much higher 
surface temperature for magnetars compared to pulsars.

%\subsubsection{The Second Stage}

{\bf The Second Stage}:
As the strong field moves through the outer crust, mechanical disturbances of the crust are 
likely to be triggered by the magnetic pressure, leading to glitches and flares. The upper crust 
would be unable to support stresses for magnetic field difference across the crust of 
$> 10^{13}$ G, as the maximum stress that the crust can support is estimated to be 
$\simeq  10^{27}$ dyne $cm^{-2}$ \cite{ruderman91}. 
Only after the core magnetic field penetrates the crust does the radiative emission and serious 
spin
down begin. The magnetars exhibit spin down ages all the way upto $10^{4(5)}$ years 
( a time scale similar to that
for ambipolar diffusion). Over this period,  the dissipation of the shielding currents, is 
responsible for their radiative luminosity and flares, till the field attains its final dipolar 
configuration.

As per the dynamo mechanism the magnetic field of a magnetar is expected to decrease with time 
from the dissipation of magnetic energy, though, there seems to be some evidence to the 
contrary \cite{thompson+02}.Our analysis of the timing data of Livingstone,Kaspi and Gavrill 
\cite{kaspiarxiv} also supports this. Such a trend is also seen in certain glitching radio 
pulsars \cite{zhang+lin05}.

Once all the screening currents dissipate away and the magnetic field reaches its fully relaxed configuration,
there is no more energy available to power the magnetar activity. This could explain the upper 
cutoff in the spin period magnetars of 
$\simeq  12$ s .

%\subsubsection{The Energy Budget}
{\bf The Energy Budget}:
From the observed energy release by magnetars we have to account for the steady  X-ray luminosity
of $10^{35}$ ergs/sec {during} the spin down age (upto $10^{4(5)}$ y.) and several flares of $10^{41 -45}$ ergs. 
Before this phase considerable energy release is likely from neutrino emission \cite{kaminker07}. A total energy in excess of $ 10^{47}$ ergs is released.

Since the shielding currents are resposible for confining the dipolar magnetic field in and around the core,
a lower bound on the amount of energy locked into the shielding currents can be estimated to be of the order of field energy of the core. For a core field of $\simeq 10^{16(17)}$ G and a core of ~ 3 km this works out  $10^{48(49)}$ ergs. 
Even if most of the energy \cite{kaminker07} is dissipated in neutrino emission during the ambipolar transport phase, we are left with enough to account for the above processes.The energy budget works out well.

That still leaves open the question of the source of energy required for setting up the shielding currents. The 
energy released in the phase transition the neutral pion condensate phase is $\simeq 30$ Mev ($4.5\cdot 10^{-5}$ ergs)  per baryon (see fig 1(b) in \cite{dbvsplb}). This energy source works out to an energy of $\simeq 10^{49(50)}$ ergs.
for a core of 3 Km of baryon density $1 fm^{-3}$.This is the driver for making the shielding currents
and this matching of energies is rather remarkable.

\section*{Observational Support}
We now examine the broad observational support that recent data has provided for our
model.
%\subsection{Progenitor Mass}
{\bf Progenitor Mass}:
Since there is a one to one correlation between the mass of the star and the central
density, in our model, only those neutron stars that exceed a certain threshold mass, $M_T$, will make  high
density magnetic cores that can lead to magnetars.
The associated progenitor mass also has to be high
enough to yield a higher mass neutron star. The density of the
magnetic core and hence its magnetic moment density and core magnetic field will all go
up as the mass, $M_T$, goes up - till, at the maximum mass, $M_{max}$ the star becomes 
unstable \cite{dbvsarxiv}.
Many authors have observed that the magnetars are descendents
of large mass progenitors or protostars. Evidence for this is given in \cite{leahy0,wickramasinghe,gaensler0,muno0}. 
{Wickramasinghe et al \cite{wickramasinghe} further suggest, based on their simulations, that magnetars
may actually belong to the higher ( than pulsars) mass ($\simeq 1.6 M_{\odot}$) population of
neutron stars.} We take these as indicators that the magnetars are more massive
than pulsars.

%\subsection{{Delayed Amplification of Surface Magnetic Field and the age anomaly}}
{\bf Delayed Amplification }:
Dar and De Rujula \cite{ddarxiv,ddgreco,dar1999a}, and Marsden et al \cite{marsden0}, find that 
the dynamo model of SGRs cannot explain their ages \cite{marsden0}.They also find that the spin down ages are much smaller than the ages of their SNR's.

The transport of the field to the surface of the star happens in two stages according to our model:
i) the transport to the crust whose time scale is
determined by ambipolar diffusion ( $10^{4(5)}$ years) and 
ii) the emergence of the field out of the crust which sets the spin down scenario for the magnetar. The first
stage is a hidden stage and is not captured by the spin down age. Thus in our model there is naturally a delayed amplification of the surface field. 

The evidence given below matches exactly with the scenario sketched out above.
Leahy and Ouyed \cite{leahy0} find that the spin-down ages of magnetars are {\em systematically} smaller 
than the ages of the associated super-nova remnants(SNR). While the spin-down ages of their sample varies between 1000-5000 yrs, the mismatch with the corresponding SNR ages is systematically of the order of $10^4 yrs.$. 
This is illustrated in Fig. 1 of their paper \cite{leahy0} which is reproduced as Fig. 2 below.
\begin{figure}[h]
\begin{centering}
\includegraphics[scale=0.60]{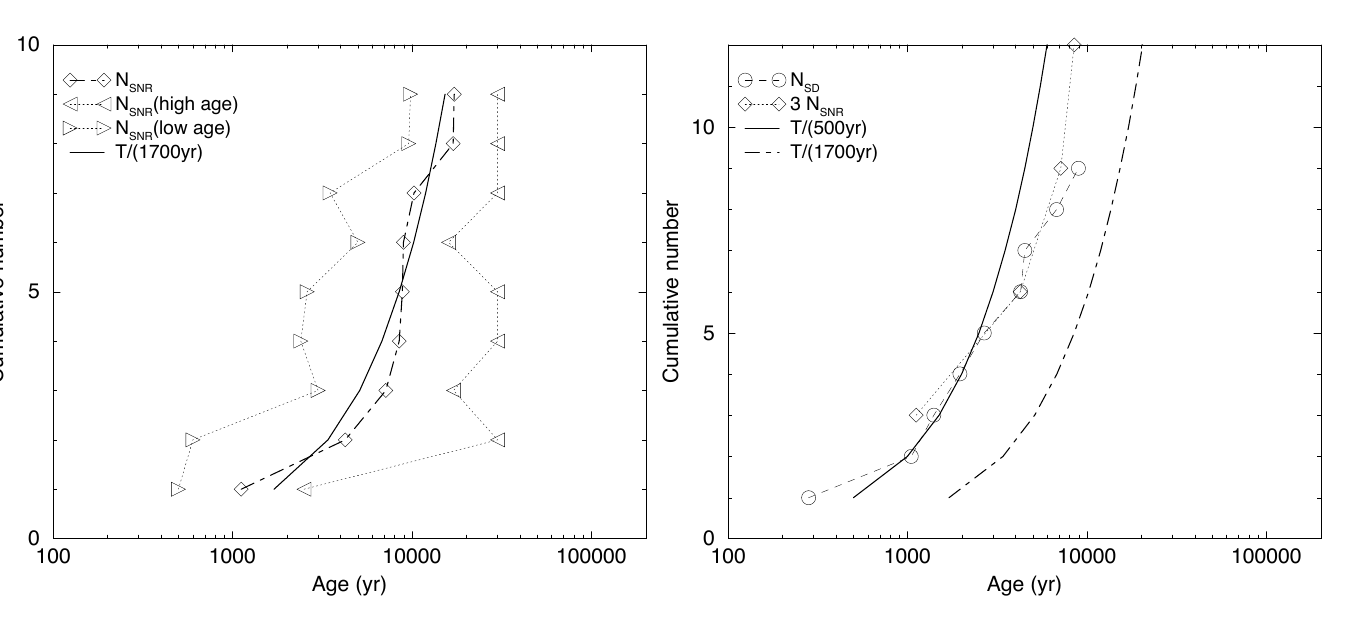}
\caption{The observational evidence for delayed amplification (From Leahy et al \cite{leahy0})}
\end{centering}
\end{figure}

The left hand panel plots the SNR ages along with a best fit curve. The right hand panel plots the
spin-down ages also along with a best fit curve. To facilitate comparison, the best fit curve for
the SNR ages is included in the right panel also. One sees a clear and systematic shift between the two best fit curves,
with the SNR's systematically {\em older}.
This evidence fits well with our picture of the delayed amplification coming from ambipolar diffusion 
including the time scale alluded to before.

%\subsection{Anomalous Spin-down: Spin down rate increases after flare as B field goes up}.
{\bf Anomalous Spin-down}: 
In the conventional dynamo mechanism the star is born with a large surface magnetic
field. Energy release at the surface then must come at the expense of the magnetic energy,
which would produce the opposite effect: reduce the magnetic field and consequently the
spin down rate.

In our model, energy is released as the crust is cleaved by the emerging magnetic field, as the shielding currents in the crust dissipate. Subsequently the core field exits to the surface increasing the surface magnetic field and the spin
down rate.

This is particularly evident in a flare phenomenon, as pointed out below by Dar and
DeRujula where the release of energy in the flare is accompanied by an accelerated spin
down rate and a consequent rise in the surface magnetic field, contrary to what the dynamo model
would predict. According to Dar and DeRujula \cite{ddarxiv,ddgreco},
in the magnetar model of SGRs the radiation-energy source is the magnetic field
energy. The spin-down rate of SGR 1900+14 roughly doubled from around the time of its
large flare on 27 August 1998 \cite{woods1999a,marsden0}.This sudden doubling of magnetic energy is what they termed
{\em energy crisis}. This has also been stressed by \cite{marsden0}.
For AXPs too the dynamo model faces similar difficulties.

These disparate features of magnetars and the fact that the progenitors have higher masses and consequently higher fields are intrinsic to our scenario. So is the delayed amplification of their magnetic fields. Further, once the core field has fully emerged after dissipation of the shielding
currents there is no further energy to release. This is the reason 
that magnetars have spin down ages between $10^4$ to $10^5$ years followed by 
cessation of activity.

\section*{Predictions}
%\subsection{Extra Neutrino emission}
{\bf Extra Neutrino emission}:
We have pointed out that during the ambipolar diffusion stage ( $10^{4(5)}$ years) most
of the energy dissipation takes place via neutrino emission. For magnetars in contrast
to pulsars, this is an additional source of neutrino emission. This would be a new observational
signature for our model.
%\subsection{New magnetars from a crossover criterion}

{\bf New magnetars from a crossover criterion}:
For older or larger period (more than 5s.) neutron stars the new feature for
magnetars is that ${\dot E}_X$, the rate of steady X-ray emission must dominate over
${\dot E}_R$, the rate for rotational energy loss from dipole radiation.

According to our model, the strong magnetic field emerges out of the
star much after its birth, unlike in the case of ordinary pulsars. 
We shall give some estimates which, though qualitative, capture some of the physics.
In our picture, when the shielding currents eventually reach the crust, the core magnetic
field is completely shielded within the crust. As the shielding currents {in the crust }dissipate, they 
restore the crust magnetic field to what its unshielded value would have been.

{ Since this field was confined by the Lenz currents in the crust, the lower 
bound on the field energy carried by the shielding currents in the crust is given roughly by,
$E_{stored} \simeq ={B_f^2}/{8\pi}\cdot 4\pi R^2\cdot \Delta R_{cr}$
where $B_f$ is the final observed surface field, $R$ the radius of the star and $\Delta R_{cr}$ the thickness 
of the crust. This  energy  will be dissipated and released as X-rays, flares etc. in a typical
time-scale of the spin-down age $\tau_{SD}$.
%However, it is not clear if
%$\tau_{SD}$ represents the true exit time. 
%The spin down age $\tau_{SD} =  {P}/{2 {\dot P}} \simeq {P^2}/{B_f^2}$ is a
%rough estimate. Depending at what stage it is used , it can
%over (under) estimate the real time taken for B field relaxation. For example,
%when the field has fully emerged or relaxed, the spin down rate ceases to
%have an increment from the incremental surface field that operated before. In
%this phase the spin down rate will slow and so will, ${\dot P}$, giving a larger spin
%down age. However, the crustal shielding currents are still about, giving a
%residual X-ray luminosity till they finally decay. We cannot tell precisely at
%what stage $\tau_{SD}$ will be a good measure of the total exit time.
Thus, an estimate for the average X-ray luminosity is given by
$$
L_X^{calc} = k\cdot R^2\Delta R_{cr}\cdot B_f^2\cdot\frac{{\dot P}}{P}
$$
where we have introduced a factor $k$ as an indicator of possible deviations from this estimate., which is a lower bound.} 

{ 
As a matter of fact we can estimate the ratio of the calculated X-ray luminosity $L_X^{calc}$ to ${\dot E}_R$. 
From our earlier discussion one sees that both these are proportional to ${\dot P}$ and hence their ratio
$$
{\cal R} = \frac{L_X^{calc}}{{\dot E}_R} = \frac{k}{4000}\cdot B_{13}^2\cdot P^2
$$
is {\em independent} of the mechanism for spin-down. In our model both $B$ and $P$ go up with time. Therefore ${\cal R}$
will keep increasing with time giving rise to a crossover between ${\cal R} < 1$ and ${\cal R} > 1$ regimes. If we use 
$k\simeq 20$ as preferred by observed data, then this ratio exceeds unity even for $B_{13}\simeq 3$ and $P\simeq 
5 s.$ For pulsars $L_X$ is absent.}

{ In table I we have summarized the observed properties of a number of high magnetic field pulsars. The magnetic fields in table are inferred from the observed $P, {\dot P}$ by taking $k=1$ and by the use of the dipole radiation formula }. According to the crossover criterion those stars marked as {\bf M} in table I are candidates for magnetars in our scenario. This crossover happens when the rotational energy loss rate (at $P \simeq 5 s.$) is too small to fuel  the X-ray flux. We therefore need another energy source for the X-radiation - in our model it is the dissipation of the shielding currents.
Pulsars do not have this source. This crossover during the evolution of a star is a signal for magnetars. We emphasize that in the conventional dynamo scenario they would not be considered as magnetars. We note that the object J1846-0258 is anomalous in many respects and needs special consideration ( see below).
%\subsection{New Young (sub second period) magnetars}
%PSR 1846-0258 has period of 326 ms. and is not generally considered to be  a
%magnetar. We shall however argue that based on our model this object should be interpreted
%as a young magnetar.

{\bf New Young (sub second period) magnetars}:
%\subsection{New Young (sub second period) magnetars}

%PSR 1846-0258 has period of 326 ms. and is not generally considered to be  a
%magnetar. We shall however argue that based on our model this object should be interpreted
%as a young magnetar.

All 'observed' magnetars have periods between $5< P< 11$ sec, magnetic 
fields between $10^{14} < B < 10^{15}$ G and a steady radiative emission of  
$\simeq 10^{35}$ ergs/sec. Why, then, are magnetars not observed 
with $ P< 5s.$ ?

In the dynamo model 
magnetars are born with high
B fields $\simeq 10^{14(15)}$ G
and would spin down very fast from periods of a few ms. to periods of over a second(in a few hundred years).
Therefore there is a much higher probability of observing them when they have slowed to periods $\simeq >$ 5 s.
and their spin down ages are much higher. In this epoch their X-ray luminosity $L_X^{obs}$ is much greater than
${\dot E}_R$. Hence it is customary to think of magnetars in the above regime. 

Observation of PSR 1846-0258 
with P=0.326 s., $B = 5\cdot 10^{13}$ G and $L_X^{obs} < {\dot E}_R$ would then rule against it being identified
as a magnetar conventionally. Furthermore, since the dipole radiation dominates over the staedy X-ray flux, the crossover
criterion discussed earlier can not be applied to it at this epoch.

The pulsar PSR 1846-0258 has a very high X-ray luminosity, comparable in magnitude to that for magnetars. 
One may then suspect that it may become a magnetar when its period evolves to $ P \simeq 5 s.$ 
We need a new signature to establish its credentials as a magnetar.
\begin{itemize}
\item Gavrill et al \cite{gavrill0}.
report a glitch accompanied by a flare in this object that are more characterstic of magnetars than pulsars.
The post flare phenomenon shows an increase in the spin
down rate.
\item A careful and  detailed timing data and analysis for this pulsar has been done by Kaspi et al \cite{kaspiarxiv}.
Part of their analysis most relevant to our model is best summarized in Fig. 4 of \cite{kaspiarxiv} which we
have reproduced in Fig. 3 here. 
\end{itemize}

\begin{figure}[h]
\begin{centering}
\includegraphics[scale=0.20]{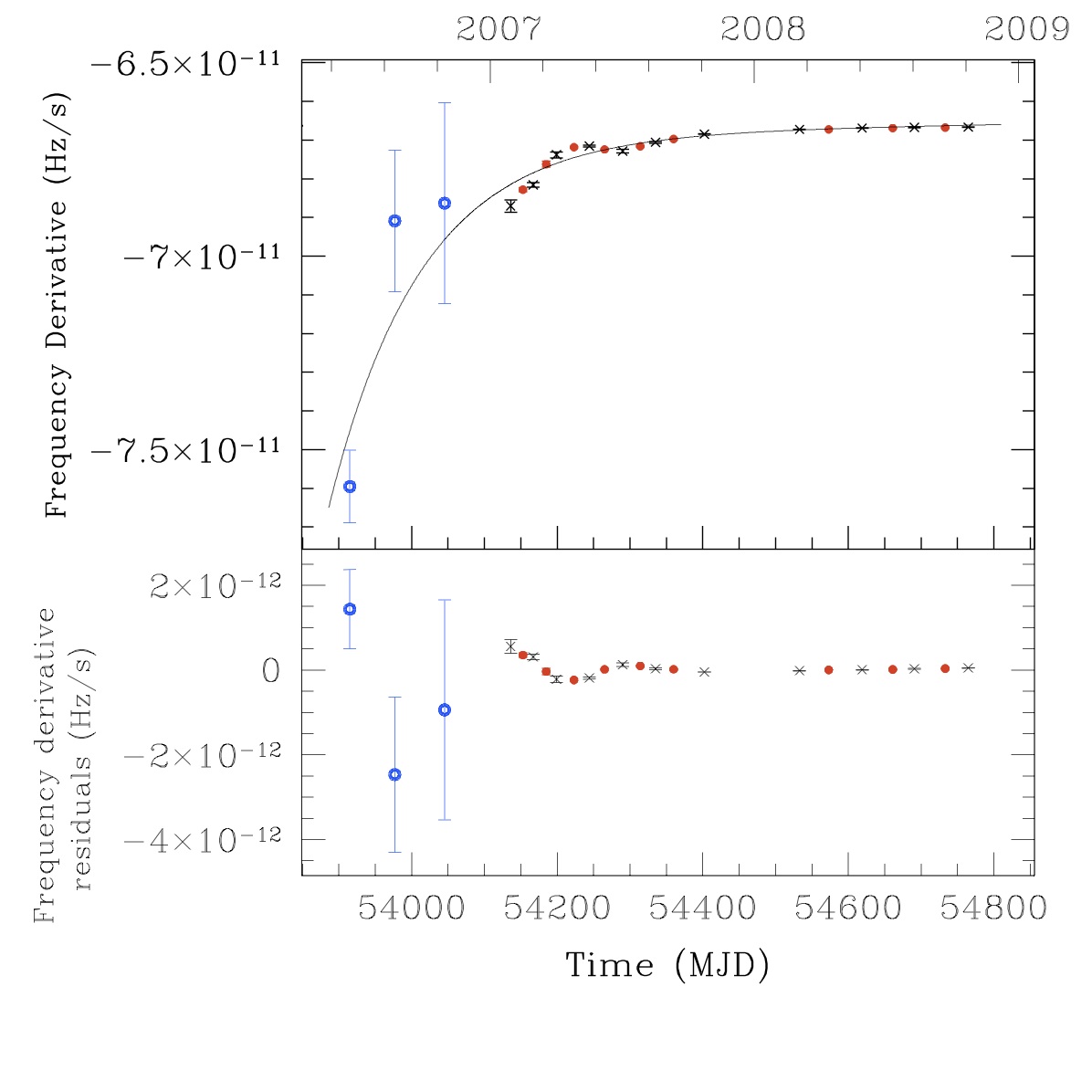}
\caption{Timing Analysis from Kaspi et al \cite{kaspiarxiv}}
\end{centering}
\end{figure}
Prior to the glitch the frequency derivative 
${\dot \nu}_0$ was $-6.3\cdot 10^{-11}s^{-2}$. 
Immediately after the glitch
this changed to $-7.55\cdot 10^{-11}s^{-2}$. They observed an exponential decay of this increase with a characterstic time
scale $\tau_d = 127 d.$ But even after 920 d.($7.25 \tau_d$) they observed an asymptotic ${\dot \nu}$ of $-6.65\cdot 10^{-11}s^{-2}$
giving a difference of $3.5\cdot 10^{-12}s^{-2}$ over the preglitch value. The exponential tail of the post-glitch increase after $7.25 \tau_d$
being $5.5\cdot 10^{-14}s^{-2}$, we interpret their data as evidence for a real change in ${\dot\nu}$ of 
$3.5\cdot 10^{-12}s^{-2}$. 
Using the relation $B = 3.2\sqrt({\dot P}~P\cdot 10^{19}$G which is based on the dipole formula
$$
{\dot P} =\frac{2}{3}\frac{4\pi^2}{P}\frac{B^2R^6}{c^3I}
$$ 
we can translate this asymptotic change in ${\dot \nu}$ to an increase in the surface magnetic field of $0.13\cdot10^{13}$ G in about 2.5 years. 
We have made use of the fact that $P$ only changed by a few ppm during this entire period.

A post flare ’ anomalous’ increment in the surface B and the spin-down rate, is predicted
by our model for magnetars. In the context of our model the anomalous behaviour of PSR
1846-0258 identifies it as a magnetar. 

It is important to see how distinctive PSR 1846-0258 is in this regard when compared to other glitching
pulsars which are obviously not magnetars. In particular, whether they too have {\em persistent} changes in ${\dot \nu}$
that are significant, in which case persistent changes in ${\dot \nu}$ would merely reflect the large glitches that took place and nothing more. We have analysed the cases of the Vela pulsar and PSR 0355+54 both of which are known to have large glitches with $\Delta\Omega/\Omega\simeq 10^{-6}$. A comparison of these two cases with that
of J1846-0258 is given in Table. 3. In the case of PSR 0355+54 \cite{alpar_pines_shaham}, Lyne's fit \cite{lyne} for
the observed post-glitch behaviour yields a persistent shift $\Delta{\dot \Omega}/{\dot \Omega} = 0.0059$
which is ten times smaller than in the case of PSR 1846-0258. In this case observations were carried out over a 77 d.
period while the characterstic timescale of the exponential decay was 44 d. In the timing analysis of Kaspi et al 
\cite{kaspiarxiv} the shift was directly observable as data was available over a much longer period.
In the case of the Vela pulsar, according to Alpar \cite{alpar_vela}, the observed long term 
breaking index of 1.4 can be explained by a persistent shift in $\Delta{\dot \Omega}/{\dot \Omega}$
of $3\times 10^{-4}$ happens about once every four years. Even this inferred effect is more than two orders of magnitude smaller than the observed effect in PSR 1846-0258. 
This indeed makes PSR 1846-0258 distinctive.  

{\bf New braking index for J1846-0258:} In a very interesting development Kaspi et al \cite{kaspibraking} report
a new braking index of 2.16$\pm$ .13 for the post-outburst phase of this object which is considerably \emph{lower}
than the pre-outburst value of 2.65$\pm$.01. They carefully analyse the sources for such a lowering of the
braking index and essentially identify three such:
\begin{itemize}
\item A \emph{positive} second time-derivative of the moment of inertia I. They conclude that it is hard to imagine
a physical situation causing this.
\item A \emph{positive} rate of change of $\alpha$, the angle between the magnetic axis and the axis of rotation.
However, according to them no change in the pulse profile has been observed over the relevant period, making this
source very unlikely.
\item Plasma distortions of magnetic fields can also be a source. But according to these authors 
that would result in a \emph{six fold} increase in particle luminosity and no such increase
has been observed.
\item An \emph{increase} in the surface magnetic field B, more precisely a \emph{positive} 
$\frac{d B}{dt}$.
\end{itemize}

Thus a persistent increase in the surface magnetic field is the most likely explanation for the
lowering of the braking index and this further corroborates our point of view that the surface magnetic field of J1846-0258 is indeed increasing and that this object is a magnetar in making.

Another possibile source for an increase in $\dot\nu$ is pulsar wind - but there is no reason to
expect it to correlate with long term post flare phenomena.

\section*{Conclusion}

In conclusion we enumerate some of the consequences of the model presented above:

i) Magnetars belong exclusively to the higher than pulsar mass population of neutron stars that have a high density magnetic core.

ii) This core is created by the strong interactions and the most likely ground state is a $ \pi_0$ condensate that aligns magnetic moments to create large dynamical $ 10^{16(17)} $ G magnetic fields at the surface of the core. Dynamical fields are 'permanent' unlike fields derived from currents.

iii) The core field is shielded by Lenz currents generated in the high conductivity plasma in and around it, but is gradually transported 
to the crust  by ambipolar diffusion over a timescale of $\simeq 10^4 $ years - this results in a time delay before the field comes out to the surface.

iv) The strong magnetic field breaks through the crust as the shielding currents dissipate giving out a steady X-ray flux and several energetic flares.

v) This further implies that the surface field keeps increasing in magnitude till all shielding currents dissipate and the permanent dipolar core field is established.

We have found that all these phenomena are supported by extensive data and observations. We have also used the model to 
identify some stars, in a sampling of a few high magmetic field pulsars, as magnetars.
Thus this model for magnetars throws up a lot of unexplored physics from the strongly interacting core to 
the plasma physics and the crustal solid state physics of huge magnetic fields. 

Our understanding of neutron stars is at a crossroad. We have to
understand many families of  neutron stars, for example pulsars and magnetars, in one
framework. This is what we have tried to do in this work. Neutron stars are also the laboratory to
understand the high density phase diagram for strong interactions. This work gives us
a new understanding of the strong interactions that is linked intimately to astrophysical
data.

\section*{Acknowledgements}
V.S. thanks Dipankar Bhattacharya for earlier collaboration that led to this work . We are
grateful to A. Deshpande and Raymond F. Sawyer for a critical reading of the manuscript and for insightful comments
that have significantly improved the paper. We thank S. Jhingan, Pankaj Sharan, C. Sivaram,
S. Mahajan, B. Paul, Deepak Kumar, P. Dasgupta, A. Khare, G. Baskaran and D. Sahdev for
discussions. Special thanks are to Siraj Hasan for facilitating initial work at the Indian Institute
of Astrophysics. V.S. thanks the University Grants Commission, National Physical Laboratory
and Centre for Theoretical Physics, Jamia Milia University for support and the CHEP, IISc,
Raman Institute and Indian Institute of Astrophysics for hospitality during the course of this
work. NDH wishes to acknowledge the Department of Atomic Energy, India, for the Raja
Ramanna Fellowship and CHEP, IISc, for the opportunity to utilise this fellowship.
\section*{Note Added}
After this work was completed and this paper written we came across a paper by Kutschera \cite{kutschera_mag} 
which has also discussed shielding and eventual outward transport of core magnetic fields. But the mechanism
discussed there is very different from Ambipolar diffusion, and that paper is addressing low field millisecond pulsars.
It estimates much larger time scales and will not be relevant for magnetars.
%We are grateful to  A. Deshpande for inputs that have significantly improved the paper. 
%V.S.  
%thanks the University Grants Commission and Centre for Theoretical Physics, 
%Jamia Milia University for support.
%NDH wishes to acknowledge the
%Department of Atomic Energy, India, for the Raja Ramanna Fellowship. Both authors thank CHEP, IISc, for hospitality.

\vspace{0.1in}
\begin{center}
{\bf TABLES}\\
\end{center}
\begin{table}[h]
\begin{center}
\caption{Compilation of actual data, $P, {\dot P}$, for high magnetic field pulsars (HMFP) from ref.\cite{gonzalez0}, with inferred 
$B,\tau_{SD},{\dot E}_R$ ( from the dipole radiation formula). $L_X^{calc}$ ( = ${E_{stored}}/{\tau_{SD}}$) is calculated 
for $k=1$. In estimating $L_X$ we have taken a star radius of 10 kms, a crust thickness of 1 km and a moment of inertia I 
of $\simeq 10^{45}~gcm^2$. Our estimates for $L_X$ are about an order of magnitude  smaller than the observed values.This 
means that the $k$-factor is in reality about 20.}
\vspace{0.1in}
\begin{tabular}{|l|l|l|l|l|l|l|l|l|}
\hline\\
%Table2
%Xray or High Magnetic Field pulsars – HMF pulsars
Name  &P  &$\dot P$ in &$\tau_{SD}$ in &$B_{13}$&${\dot E}_R$ in & $L_X^{obs}$ in & $L_X^{calc}$ in & $L_X^{obs}/L_X^{calc}$   \\
 & in sec. & $10^{-12}$ & $10^3$ yrs & & $10^{33}$ ergs/s&  $10^{33}$ ergs/s& $10^{33}$ ergs/s &    \\
\hline
J1718-3718 {\bf M}& 3.3& 1.5& 34&7.4& 1.6& 6&0.25 &24\\
\hline
J1846-0258 {\bf M} & 0.33& 7& .7& 5& 8300& 100 &5 &18\\
\hline
J1119-6127& 0.407& 4.1& 1.6& 4.1& 2600& 0.216&1.7 &0.13\\
\hline
J1847-0130 {\bf M}& 6.7& 1.3& 83& 9.4& 0.17& 3 to 8 &0.2 &19\\
\hline
J1814-1744 {\bf M} & 4.0& 0.74& 85& 5.5& 0.47& 4.3 &0.06 &72\\
\hline
PSR0154-61& 2.35& 0.189& 197& 2.1& 0.57& 0.14 &0.004 &35\\
\hline
%RXJ 0007.0 +7303  & 0.316& 0.361& 14& 2.1& 160&  & &\\
%\hline
\end{tabular}
\end{center}
\end{table}
%
%\vspace{0.1in}
\begin{table}[h]
\begin{center}
\caption{ A comparison between pulsars at the beginning of spin-down, high magnetic field pulsars (HMFP)and typical magnetars. 
Other parameters are calculated using the dipole formula with $P, B$ as input.In the first row we have given the typical values 
of  $P, B$ at the beginning of the spin down phase.The second row is constructed by taking the average values of $P, B$ from 
table I (HMFP)for all cases marked {\bf M} (except J1846-0258). Finally, the last row is obtained by taking for $P, B$ from 
the average of the AXP data given in Gonzalez et al \cite{gonzalez0}.}
%
%
%\vspace{0.1in}
%
\begin{tabular}{|l|l|l|l|l|l|l|l|}
\hline\\
Name      & P & $\dot P$ & $\tau_{SD}={P}/{2\dot P}$ &$E_R$ &${\dot E}_R$  &$B_{13}$& $L_X^{calc}$ \\
&  in sec.&  in $10^{-12}$&  in $10^{3}$ yrs.& in ergs.& in ergs/s&  &in ergs/s\\
\hline\\
Beginning of spin down& 0.1& 1.0& 1.666  &$2\cdot 10^{48}$& $4\cdot 10^{37}$&  1.0& $ 10^{32}$ \\
\hline
Typical High Mag Field HMF & 5.0& 1.13& 74.0 & $8\cdot 10^{44}$& $3.6\cdot 10^{32}$&  7.5& $10^{32}$\\
pulsar or magnetar: $P>5$s& & & &   & & &\\
\hline
Typical magnetar $P>5$s& 9.0& 2.5& 60& $2.5\cdot 10^ {44}$& $1.25\cdot 10^{33}$&  50.0& $6\cdot 10^{34}$\\
\hline
%$R_{2b}$ &  Expected &   Predominant & 1/5-1/7 with 100\% eff& 1/12& \\
%\hline
\end{tabular}
\end{center}
\end{table}
%\vspace{0.1in}
\begin{table}[h]
\begin{center}
\caption{Comparison of persistent $\Delta{\dot\Omega}/{\dot\Omega}$ of PSR 1846-0528 with those in some large-glitch pulsars.}
%\vspace{0.1in}
%
\begin{tabular}{|l|l|l|l|l|}
\hline\\
Name          &Length of    &Exponential          &Persistent                         &Source  \\
              & observation & decay time $\tau_d$ & $\Delta{\dot\Omega}/{\dot\Omega}$ &   \\
\hline\\
PSR 1846-0528 & 920 d.      & 127 d.              & $5.5\cdot 10^{-2}$                & Timing Data \cite{kaspiarxiv} \\
\hline
PSR 0355+54 & 77 d. & 44 d. & $6\cdot 10^{-3}$ & Timing Data \cite{lyne}\\
\hline
Vela & -- & -- & $3\cdot 10^{-4}$ & Breaking Index \cite{alpar_vela}\\
\hline
\end{tabular}
\end{center}
\end{table}
\end{document}